\documentstyle[11pt,aaspp4,psfig]{article}

\def\5ov8{\textstyle {5 \over 8}   \displaystyle}
\def\fourth{\textstyle {1 \over 4} \displaystyle}
\def\unvc{ {\mbox{\boldmath $1$}} }

\def\itl{\'{\i}}

\def\etal{{\it et al.~}}

\begin{document}
\newcommand{\figureout}[4]{\psfig{figure=#1,width=#2,angle=#3} 
   \figcaption{#4} }

\baselineskip 0.7truecm

\title{\bf Cyclotron Line Features from Near-Critical Fields II: \\
		 on the Effect of Anisotropic Radiation Fields}

\author{Rafael A. Araya-G\'ochez}
\affil{Laboratorio de Investigaciones Astrof{\itl}sicas, \\
	Escuela de F{\itl}sica, Universidad de Costa Rica, \\
	San Jos\'e, Costa Rica.} 
\and 
\author{Alice K. Harding}
\affil{
 Laboratory for High Energy Astrophysics, \\
 NASA Goddard Space Fight Center, Greenbelt, MD 20771.}


%--\date{}
%--\maketitle

\begin{abstract}  
 We assess the impact of radiation anisotropy on the 
 line shapes that result from relativistic magnetic
 Compton scattering in the low-density/high-field regime.  
 A Monte Carlo implementation of radiation transport allows
 for spatial diffusion of photons in arbitrary geometries
 and accounts for 
 relativistic angular redistribution. %-- (Araya \& Harding 1999). 
 The cross section includes natural line widths and 
 photon ``spawning" from up to fourth harmonic photons. 
 In our first paper we noted that even if the photon injection 
 is isotropic a strongly anisotropic radiation field rapidly 
 ensues.  %-- even at relatively low optical depths.  
 We now investigate the angular distribution of cyclotron spectra 
 emerging from an internally irradiated magnetized plasma with
 a prescribed global geometry (either cylindrical or plane parallel)
 and the effects of anisotropic photon injection on the line shapes.
 Varying the input angular distribution permits a better 
 understanding of the line formation process in more realistic
 scenarios where the radiative mechanisms are influenced by the
 intrinsic anisotropy of the field and by moderate relativistic 
 beaming. In general, the line features are
 most pronounced along the directions of the anisotropic continuum 
 injection and tend to be weakened in other directions, relative 
 to the line features resulting from an isotropic continuum injection. 
 We find that the enhancements at the line wings of the fundamental, 
 which appear prominantly 
 in the case of isotropic continuum injection, are strongly suppressed 
 along the direction of anisotropy in the case
 of beamed continuum injection, regardless of geometry or beaming pattern.
\end{abstract}    

%###########################################################################

\section{INTRODUCTION}            
 \label{sec:Intro}
    
       The most direct observational evidence of the magnetic
 field strength in neutron star sources is provided by X-ray
 cyclotron line features.  Such quasi-harmonic spectral features
 are generated by scattering of resonant photons with the
 electrons embedded in the magnetosphere.  
 The interaction is multiply resonant because the Dirac equation 
 in the presence of a locally uniform magnetic field, {\bf B}, 
 yields discrete perpendicular momentum eigenvalues 
 (with $c = \hbar \equiv 1$): 
 $(p_{\perp}/m_e)^2 = 2 n(B/B_{\rm Q})$, %--(e.g. Melrose and Parle 1983), 
 but relativistic effects introduce 
 a slight anharmonicity in the rest-frame resonant photon energies: 
\begin{equation} 
 	{\omega}_n/m_e = [(1 + 2n B'{\rm sin}^2\theta)^{1/2} - 1] 
		/ {\rm sin}^2 \theta, 
\label{eq:NResEner} \end{equation} 
 where $B' \equiv B/B_{\rm Q}$ and 
 $B_{\rm Q} = m^2/e (c^3/\hbar) \simeq 44.$ TG  
 is the Q.E.D. field scale.  
 As these energies depend non-trivially on the photon's 
 propagation angle with respect to the field, 
 $\theta = \cos^{-1}(\hat{\bf k}.\hat{\rm \bf B})$, 
 the emergent spectral features are heavily influenced by 
 the angular distribution of resonant photons as well as by 
 the spatial distribution of electron targets 
 (i.e. plasma ``geometry") and by the scattering 
 region's orientation with respect to the magnetic field.

        Cyclotron lines have been detected in a number of X-ray 
 pulsar spectra (Nagase 1989, Mihara, 1995, Santangelo et al. 1999) 
 and possibly also in the spectra of several Gamma Ray Bursts
 (Mazets \etal 1981, 1883 Murakami \etal 1988, Yoshida \etal 1991). 
 While there is a broad consensus on the origin of the former
 objects, the origin of gamma ray bursts remains highly elusive.  
 Line energies in X-ray pulsars typically fall in the range
 from 4 to 40 keV, 
indicating\footnote{
 Note that the standard interpetation of field strength 
 through the cyclotron fundamental 
 $\omega_{\rm cyc} = m_e (B/B_{\rm Q})$ assumes a hydrostatic plasma
 and is not unique.  Models assuming bulk motion of the scattering 
 electrons (e.g. Baushev \& Bisnovatyi-Kogan 1999) will produce a
 Doppler-shifted line center, but the narrowness of the lines 
 is then difficult to accomodate.} 
 scattering of electrons in magnetic
 fields of up to $3.5$ TG.   The case of transient X-ray pulsar 
 A0535+26 with an 11 TG field ($\sim \fourth B_{\rm Q}$) represents
 the uppermost limit spectrally determined (Grove \etal 1995, 
 Araya and Harding 1996a-b, Maisack \etal 1997)
%-- and independently confirmed (Bildsten \etal 1997).  
 
	X-ray pulsars are found in binary systems containing a 
 neutron star.  In high mass X-ray binaries (HMXRB), a tenuous 
 accretion wind occurs from the massive companion onto the neutron  
 star surface.  Low mass systems have higher accretion rates and 
 disks surrounding the neutron star (see van den Heuvel \& Rappaport
 1987, for a review).  The plasma geometry depends on the mechanism
 for stopping the accretion flow at the surface of the star, 
 which is not certain in either case.
 If shocks form due to radiation pressure 
 in high luminosity sources (Basko \& Sunyaev 1976, Davidson 1973, 
 Wang and Frank 1982, Arons, Klein and Lea 1985, 1987) or 
 through a collisionless plasma instability in the flow of 
 low luminosity sources (Langer and Rappaport 1982), the 
 geometry of the X-ray emitting region will be  
 a quasi-cylindrical column.  
 This situation sharply contrasts the case of flow stopping 
 through Coulomb collisions with the ambient electrons and protons where 
 atmospheres with scale heights of order $h \sim$ 200 cm are expected
 (M\'esz\'aros \etal 1983, Harding \etal 1984).  
 These possible pictures of the flow provide two idealized, 
 `orthogonal' geometric scenarios: plane parallel slabs and 
 cylinders.  One hopes that models of cyclotron line 
 features may provide a sensible diagnostic to determine the 
 geometry of the region where the lines form.  This, in turn, 
 could also help answer some important questions regarding the 
 microphysics involved in the stopping and cooling of the flow.

	Unfortunately, 
 the production of theoretical cyclotron line features in an 
 entirely self-consistent manner is a formidable task requiring
 the resolution of a coupled radiation-magnetohydrodynamic system 
 (e.g., Arons, Klein and Lea 1987).  
 Even simplified approaches pose extremely difficult problems
 so the solutions are typically attempted in piecemeal form.  
 For cyclotron features produced in hard incident continua 
 (as in this investigation) the work of Isenberg, Lamb and Wang
 (1997, Lamb \etal 1989, Wang \etal 1989, Wang, Wasserman and 
 Salpeter 1988) and of Alexander and M\'esz\'aros (1989, Alexander,
 M\'esz\'aros and Bussard 1989) represent the state of the art 
 calculations and are closely related to the models produced
 here (for a comparison, see Araya \& Harding 1999 [hereafter AH99]).

	Our results ensue from a hard incident continuum 
 (photon number power index = -1) throughout the entire 
 energy range.  This is motivated mainly by GRB spectra although 
 some X-ray pulsars exhibit a fairly hard X-ray continuum at the 
 characteristic cyclotron energy (i.e. X0115+634, X0331+53, 
 4U 1538-52 and even A0535+26 at 110 keV). 
 A hard incident spectrum allows for an easier interpretation
 of several processes affecting the formation of the fundamental 
 cyclotron line feature such as relativistic angular 
 redistribution of resonant photons and photon spawning in which 
 electrons excited to higher Landau states produce additional 
 photons through cyclotron decay.  The theoretical spectra thus
 generated greatly aid in the interpretation of real spectra 
 formed in softer continua such as those belonging to some 
 X-ray pulsars (where the effect of photon spawning is 
 substantially weaker).

 	In AH99, we presented model cyclotron line features for
 isotropic photon continuum injection.  It was noted then that 
 radiation transport and resonant scattering efficiently 
 induced a strongly anisotropic radiaton field.  Moreover,  
 the radiation processes in very strong magnetic fields and
 the inherent anisotropy induced by the plasma geometry are
 unlikely to produce isotropic continuum emission except 
 perhaps deep inside the accretion column or mound.
 In this paper, we offer a detail quantification of the emergent
 angular distribution for isotropically injected photons
 scattered off electrons with a thermal energy 
 distribution parallel to the field,
 and expand the calculations to encompass the influence of 
 anisotropic photon injection on the emergent cyclotron line
 features. 
%-- Electrons in the plasma are assumed to have a thermal 
%-- distribution parallel to the field.

	In \S \ref{sec:ModDesc} we present a synopsis of the
 model and %--physical assumptions that delimit
 its regime of validity.
 Motivation to the study of anisotropic photon injection and details
 of its implementation are presented in \S \ref{sec:Motiv}  
 This scheme is then employed to study the line formation process, 
 the emergent spectral line shapes and the angular distribution 
 of emergent radiation in \S \ref{sec:ResDisc}

%#######################################################################
\section{Model Synopsis} 
 \label{sec:ModDesc}

 	The cyclotron line spectra are produced in {\em irradiated}
 magnetized plasmas without internal photon sources and with 
 {\em prescribed} thermodynamic conditions.   
 The chief simplification in our treatment of the radiation 
 transport problem is conformity with the high-field/low density 
 regime.  Vacuum polarization then determines the photon states,
 and collisional interactions may be ignored.  We have quantified
 this and the statements below in AH99.

	The relativistic magnetic Compton scattering cross section 
 (Sina 1996, see AH99) includes natural line widths through the use
 of dressed electron propagators evaluated in the presence of a uniform
 magnetic field (Graziani 1993, Graziani, Harding \& Sina 1996).
 The electron states that permit such construction are simultaneous
 eigenstates of the {\it Sokolov-Ternov} (1967) relativistic 
 spin operator (see, e.g., Melrose \& Parle 1983, Sina 1996) 
 while the photon polarization states are the magnetized 
 vacuum normal modes uncorrected for dispersion (Shabad 1976).

 	In the strong field limit, use of the normal photon modes
 yields mode-dependent cross sections which are not substantially
 different from one another and mode switching branching ratios
 strongly couple the two modes.  Wang, Wasserman and Salpeter
  (1988) thus found that both normal modes, 
 $\mbox{\boldmath $\epsilon$}_\parallel$ and 
 $\mbox{\boldmath $\epsilon$}_\perp$, have comparable mean
 free paths and contribute about equally to the spectral 
 angular output distributions in slab geometries.  
 This motivates the treatment of unpolarized photons.
 
	Detailed consideration of `Landau-Raman' scattering 
 allows photon spawning from up to fourth harmonic photons with
 electron momenta sampled from a relativistic thermal energy 
 distribution function %-- where the electron momenta corresponds
%-- to a one dimensional, `parallel' temperature:   
\[ f_e(p)dp = N_{T_e}
             \exp \left\{ -{(\sqrt{1+p_e^2}-1) \over T_e} \right\}dp
\] where $T_e$ is the (one dimensional) parallel electron temperature, 
 and $N_{T_e}$, the normalizaton.
 Regarding the Landau level populations, one need not
 consider detailed equilibrium in plasmas which are tenuous
 and cool relative to the cyclotron energy:
 For sub-critical fields, the radiative cyclotron 
 de-excitation rate (Latal 1986),
 $\nu_r \simeq  2._{+15} B_{12}^2 \, {\rm sec}^{-1}$,
 is very large compared with the collisional rate 
 (Bonazzola, Heyvaerts and Puget 1979),
 $\nu_c =  5._{+8}~(n_e/1.0_{+21}{\rm cm^{-3}})
	\, B_{12}^{-3/2} \, {\rm sec}^{-1}$.
 Thus, all electrons may be safely assumed to be in the 
 ground Landau level initially.
 
	In spite of the tenuous environment,
 a thermal characterization is justified by the resonant
 behavior of the magnetic Compton scattering cross section 
 which ensures high optical depth in the line core 
 and strong photon-electron coupling through multiple resonant
 scattering.
 An estimate for the equilibrium electron temperature,
 obtained from non-relativistic numerical simulations by 
 Lamb, Wang and Wasserman (1990), also yields consistent results
 for hard continua (AH99):
 $T_e \simeq \fourth\omega_{\rm cyc},~\alpha_{\rm eff} \sim -1,~\tau \gg
1,$
 where $\omega_{\rm cyc} = m_e (B/B_{\rm Q})$ 
 is the nominal cyclotron frequency and $\alpha_{\rm eff}$
 the effective power law index at the line.  Thus, in our calculations
 we take $T_e = 5.1$ keV for $B' = 0.04$,  $T_e = 12.8$ keV for $B' = 0.1$,
 $T_e = 31$ keV for $B' = 0.24$.

	The Monte Carlo implementation for photon injection, 
 propagation, and scattering permits the spatial diffusion of 
 photons from arbitrary photon source functions and in arbitrary 
 plasma geometries.  A relativistic photon angle redistribution 
 function, accurate for core scattering events that leave the 
 electron back in the ground level and approximate otherwise, 
 is also used to allow for angular diffusion. 

	The ``absorption" approximation used to sample angular 
 redistribution and to choose the target electrons' initial 
 momenta in the
majority\footnote{except at the last scattering event 
 where the program makes use of more accurate algorithms 
 to ``double-check" that the photon should indeed escape.}
 of scattering events is strictly valid as long as the resonant 
 part of the cross section dominates.   Accuracy in the line 
 core is warranted but not in the wings where the full scattering 
 cross section provides a better description.

	Each spectrum is formed by unbiased injection of fifty
 thousand individual photons from $N_E$ identical energy bands 
 evenly distributed throughout the spectral range.  This is
 followed by re-processing of each photon through scattering 
 and subsequent collection of the escaping radiation in a similar 
 energy grid.   To avoid poor statistics at high energies, photons
 are injected from a flat spectrum and are assigned spectral weights.  
%-- For isotropic 
%-- injection a lower limit on the optical depth at line center is 
%-- $\tau > \tau_{min} = -\ln ( {N_\mu N_E}/{N_{inj}} ) \simeq 5$,
%-- where $N_E = 80$ energy bands, $N_\mu = 4$ angle bins and 
%-- $N_{inj} = 5._{+4}$ injected photons 
%-- (note the utilization of a very compact number notation: 
%-- $5._{+4} \equiv 5.0 \times 10^{+4}$).

 	Two geometries for the scattering region are considered:
 plane parallel slab (with the magnetic field parallel to the 
 slab's normal vector) and cylindrical (with the field parallel
 to the cylinder axis).   Because of axial symmetry, the 
 $\hat{\bf 1}_z$ direction is oriented along the magnetic field 
 {\bf B}.  A continuum photon spectrum is incident from a source 
 at the slab midplane or from the cylinder axis.  Internally, the 
 source code keeps track of $N_\mu = 20$ equally spaced values of 
 $\mu \equiv \hat{k}.{\rm\bf B} \in [-1,~1]$, the cosine of the 
 viewing angle to the field, and of $N_E = 80$ energy bins with 
 $k^0 \in [\omega_{min},~\omega_{max}]$.  The injected and escaping 
 photons are accumulated into four ranges of $\mu \in [0,1]$, and 
 angle-dependent spectra are formed by considering the number of 
 emerging photons in each of the $N_E$ energy bins.  The Thomson
 optical depth of the plasma in the line forming region, 
 the electron temperature and the spectral index 
 of injected photons are prescribed.

%###########################################################################

%    Resonant Cyclotron Features from Near-Critical Magnetic fields,
%               V.  Gamma Ray Burst Type Spectra
%_________________________SECTION OUTLINE__________________________________
%~~~~~~~~~~~~~~~~~~~~~~~~~~~~~~~~~~~~~~~~~~~~~~~~~~~~~~~~~~~~~~~~~~~~~~~~~~
%  \section{Introduction}                                             L020
%  \section{Model Parameter Selection}                                L081
%  \section{Optical Depth and Geometry Variation}                     L136
%  \section{Non-Isotropic Photon Injection}                           L326
%    {fig:Profileb.04}  {fig:Profileb.24}                             L504
%    {fig:SlbOpDepth}   {fig:CylOpDepth})                             
%    {fig:Is.v.Cob.04}  {fig:AngDistb.04C}                            L576
%    {fig:Is.v.Fab.04}  {fig:AngDistb.04F}                            L60
%    {fig:Is.v.Cob.1}   {fig:AngDistb.10C}                            
%    {fig:Is.v.Fab.1}   {fig:AngDistb.10F}                            
%    {fig:Is.v.Cob.24}  {fig:AngDistb.24C}                            
%    {fig:Is.v.Fab.24}  {fig:AngDistb.24F}                            
%##########################################################################

\section{Anisotropic Photon Injection} 
 \label{sec:Motiv}

\subsection{Motivation}

     Our primary motivation for investigating the effect of 
 anisotropic continuum photon distributions on cyclotron line 
 formation is the fact that almost all radiation processes in
 the strong magnetic fields of X-ray pulsars produce anisotropic
 radiation.  Such anisotropy results from the strong angle dependence
 of the magnetized cross sections for Compton scattering and 
 radiation rates for cyclotron and bremsstrahlung (e.g. Meszaros 1992).
 A continuum spectrum resulting from inverse Compton scattering 
 would have a flux peak along the magnetic field direction, while 
 for cyclotron emission from non-relativstic electrons the radiation
 peaks perpendicular to the field direction.  

     In addition to the
 expected anisotropy of continuum radiation processes, the cyclotron
 scattering process itself produces anisotropies through angular
 redistribution.  In AH99, we found that isotropically injected 
 photons scattering through a slab were preferentially
 beamed along the magnetic field direction normal to the slab.  
 Meanwhile, isotropically injected photons scattering through a
 cylinder were beamed across the field direction.  Thus, it is of
 intrinsic interest to explore the interplay between continuum
 anisotropy and scattering-produced anisotropy.
 
 	Since the opening angle for the injection is very small
 (see Eq [\ref{eq:AngInjForm}]), the injection imitates the behavior of a 
 $\delta$-function for injection at angles parallel and perpendicular 
 to the {\bf B}-field.  
 The study of beamed injection in these two limiting cases will help to 
 understand, interpret and `de-convolve' the linear coupling of the 
 cyclotron line shapes to the anisotropies of the radiation field.

     The essential features of cyclotron line formation in
 the high-field/low density regime were discussed in AH99.
 In that paper we commented on the effect of increasing the 
 magnetic field strength from $\approx .04$B$_{\rm Q}$ to 
 $\approx .24$B$_{\rm Q}$.  Because the nominal estimate of 
 electron temperature scales directly with field strength, 
 the features at near-critical fields generally become very broad. 
 Moreover, although there may have been some expectation that at 
 large fields low harmonic features would resemble the fundamental
 line, this was not born out in our numerical results because the 
 angular redistribution function is inherently different for photons 
 at the fundamental cyclotron energy.  At low fields, $n \neq 1$ 
 cyclotron line features are `absorption-like' because the electrons 
 prefer to de-excite in single perpendicular momentum quanta thus 
 spawning photons just short of the fundamental cyclotron energy. 
 The dynamics of line formation is thus most complicated at the 
 fundamental energy.  In hard continua, photon spawning produces 
 very significant, broad line wings which distinguishes the 
 fundamental feature from all others.

				%-- angle dependence and slab vs cyclinder

\subsection{Model Parameter Selection}

 	All of the runs have a continuum optical depth 
 $\tau_c = 1._{-3}$.   The incident photon number spectra depend 
 on energy through a simple power-law 
\begin{equation}
 \frac{dN_\omega}{d\omega d\mu} = \Theta(\theta) \times \omega^\alpha 
				  ~~{\rm with}~~ \alpha = -1.
\label{eq:HardSpec} \end{equation}
 The angular dependence of the injected spectra, $\Theta(\mu)$,
 models three limiting cases: isotropic, cone beam 
 (where individual photons are injected along the magnetic field) 
 and fan beam (with photons injected perpendicular to the 
 {\bf B}-field).  We choose a Lorentzian function 
 for the probability distribution of angular injection of photons
 because it allows a simple implementation of the anisotropic 
 injection process through Monte Carlo sampling.
 The functional form of the angular distribution is
\begin{equation}
 \Theta(\theta) = \frac{1}{\pi} \left( \frac { (2/\theta_{op}) }
          {\left[ {2\over\theta_{op}} (\theta-\theta_0) \right]^2 + 1 }
 				\right),
\label{eq:AngInjForm} \end{equation}
 where: $\theta_{op} = 10^0$ is the opening angle, or `half-width' 
 of the cone or fan, and where $\theta_0$ is the center of the
 distribution: $\theta_0 = 0$ for the cone and 
 $\theta_0 = \pi/2$ for the fan distribution. 
%--the Lorentzian function for 

%##########################################################################
\section{Results and Discussion}
 \label{sec:ResDisc}

%-- Comment: what about a table for \Re ? 

	As we have already emphasized, the angular dependence of
 the emerging photons in both geometries show noticeable departures
 from isotropy even when the injection is done isotropically 
 (examine the top graphs in Figs.  
 \ref{fig:AngDistb.04C}, \ref{fig:AngDistb.04F},  
 \ref{fig:AngDistb.10C}, \ref{fig:AngDistb.10F}, 
 \ref{fig:AngDistb.24C}, and \ref{fig:AngDistb.24F}).   
 The most plausible causes for this are
 the spatial bias induced by the geometry of the  
 scattering region and the relativistic forward beaming of scattered 
 photons in the direction of the intermediate electron's momentum 
 ${\bf p}_e$ (i.e. parallel or anti-parallel to the {\bf B}-field). 
 As expected, slab geometries show the strongest anisotropy: 
 The effect of forward scattering (along the field) added to 
 the most favorable photon escape conditions at small angles 
 to the field account for this result. 

 	As a rough measure of anisotropy we define a benchmark ratio:
 the escaping photon number flux @ $\mu \simeq 1$, 
 to photon number flux @ $\mu \simeq 0$,
 $\Re^{\rm Injec.bias}_{\rm Geometry}$,
 with a super-index indicating the injection bias 
 (C for cone, F for fan and I for isotropic)
 and a sub-index indicating geometry 
 (sl for slab and cy for cylinder). 

 	For plane parallel slabs and isotropic injection at the lower 
 field strengths (B'=.04 and $\tau_c = 1._{-3}$), 
 the ratio of photon number flux in the 
 $\unvc_z$ direction to the flux perpendicular to $\unvc_z$ is 
 $\Re^{\rm I}_{\rm sl} = 3.8$.  For the cylinder, this same ratio is 
 $\Re^{\rm I}_{\rm cy} = .60$ and the radiation escapes preferentially
 in the direction of least optical depth ($\mu \rightarrow 0$). 
 
 	The anisotropy of the escaping radiation increases with 
 optical depth (Araya 1996).  For the slab (@ B'=.04 and $\tau_c =
3._{-3}$), 
 $\Re^{\rm I}_{\rm sl} = 6.0$, but remains about the same for the cylinder,
 $\Re^{\rm I}_{\rm cy} = .64$, since the depth in the parallel direction
 is already large for this geometry. 

	In larger fields, the emerging photon anisotropy for slab 
 remains about constant ($@ \tau_c = 1._{-3}$, B' = 0.24); 
 $\Re^{\rm I}_{\rm sl} = 3.3$, and decreases slightly for the cylinder:
 $\Re^{\rm I}_{\rm cy} =  .46$.  This may seem surprising at first: 
 For $B' = .24$ and $T_e = \fourth B' = 31$ keV, one would expect stronger
 forward beaming from the mildly relativistic electrons of the plasma. 
 However, the resonance profiles are stronger at lower fields (AH99).  
 This seems to play a more dominant role in the angular biasing than does
 the relativistic forward beaming in this parameter space.

      Let us examine next the effect of {\bf beamed photon injection}  
 (all fluxes will now correspond to a continuum optical depth 
 $\tau_c = 1._{-3}$).  For {\it cone} injection at the lower fields
 ($B' = .04$) our benchmark ratios are
 (Fig.\ref{fig:AngDistb.04C}):
 $\Re^{\rm C}_{\rm sl} = 1.2_{+3}$ and  
 $\Re^{\rm C}_{\rm cy} = 1.7_{+2}$,
 and for fan beam injection (Fig.\ref{fig:AngDistb.04F}):
 $\Re^{\rm F}_{\rm sl} = 0.76$ and 
 $\Re^{\rm F}_{\rm cy} = 0.12$.  Clearly,
 cylinder geometries hinder large numbers of photons from emerging at small
angles
 even if all the photons are injected there (cone injection) and the same
can
 be said for spectra seen at large angles from slabs with fan beam
injection.

 	However an unexpected result is now apparent;
 the scattering of photons injected perpendicular to the field tends to
 make these radiation field ratios more isotropic while the opposite
 happens for
 cone injection (note that the benchmark ratios are a {\bf rough} measure
of 
 isotropy; for a better description refer to the
 figures for the exact angular dependence of the emergent radiation).  
 If relativistic scattering along the field direction is the cause of 
 this behavior, then this trend 
 should be stronger for the higher fields.  

	The ratios for parallel to
 perpendicular photon number flux at higher fields are:
 $\Re^{\rm C}_{\rm sl}(.10) = 1.4_{2},~~\Re^{\rm C}_{\rm cy}(.10) = 6.8$
 (Fig.\ref{fig:AngDistb.10C}) and
 $~~\Re^{\rm C}_{\rm sl}(.24) = 1.1_{2}, ~~\Re^{\rm C}_{\rm cy}(.24) = 4.2$
 (Fig.\ref{fig:AngDistb.24C}),
 for cone injection and
 $~~\Re^{\rm F}_{\rm sl}(.10) = .48, ~~\Re^{\rm F}_{\rm cy}(.10) =
4.5_{-2}$
 (Fig.\ref{fig:AngDistb.10F}) and
 $~~Re^{\rm F}_{\rm sl}(.24) = .47, ~~\Re^{\rm F}_{\rm cy}(.24) = 4.3_{-2}$
 (Fig.\ref{fig:AngDistb.24F}), for fan injection 
Thus, note that our conjecture
 regarding the effect of relativistic scattering proves to be wrong: 
 Fan beam injection at larger fields yield emerging spectra that are beamed
along
 the perpendicular direction to the field, while the emerging radiation
 from cone injection is more isotropic!   

	Again we turn to the line profiles for an explanation.  
 The fact that the total line profiles are higher at lower fields partially
 explains the observed behavior.  Moreover, one may now attempt to find 
 further causes for the hindering of the relativistic beaming.  As was 
 pointed out by Lamb et al (1989) for low fields in the 
 non-relativistic limit, photons are removed from the 
 incident spectrum at higher harmonic energies (see the
 branching ratios in AH99) at rates $\propto sin^{2(n-1)}\theta$ and the
 spawned photon contribution is injected preferentially along the field
 $\propto 1 + \cos^2\theta$.  This trend for angular redistribution
 ({\it modulo} relativistic corrections) is borne out qualitatively
 in our low field models.  In higher fields the higher harmonic 
 photons are not removed from the spectrum but are redistributed with
similar 
 angular dependencies (as derived from the cross sections in the rest-frame
 of the initial electron and in the `$0p$-frame' of the intermediate 
 electron (see Eqs: [57] and [59] of AH99), due to the preference for $0 
 \rightarrow n \rightarrow 0$ scattering.  
%-- dominance of -for- preference for

	The emergent spectra are shown in Figs. 1, 3, 5, 7, 9, and 11.  
 Because the spectra are constrained by limited statistics, photon
 depletion at line center is unavoidable with the limited samples.
 Crosses indicate parts of the spectrum where photon depletion has
occurred. 

 	First, observe that in order to obtain an absorption
 dip for the first harmonic, the observer must see the spectra
 at an angle where injection has taken place: along the 
 {\bf B}-field for cone injection or perpendicular to {\bf B} 
 for fan injection.  
 Moreover, beginning with Fig. [\ref{fig:Is.v.Cob.04}], 
 notice that by injecting the photons parallel
 to the field (cone injection) in a slab geometry, 
 the broad line wings that occurred at small angles in the
 isotropic injection case disappear.  Instead, the feature at 
 the fundamental resonant energy is absorption-like for small
 angles and at large angles it yields an emission line! 
 (somewhat resembling the case of isotropic-injection in 
 cylindrical-geometry).
 In addition note that for a cone injected slab the features at large 
 angles are produced mostly by photons scattered out of small angles and 
 that there is no replenishing of these photons, so that exaggerated
 features are expected.  
 Thus, {\it the disappearance of the broad line wings for 
 $\mu \sim 1$ suggests that these are formed by angular
 redistribution from $\mu \sim 0$}.

    Recall from our discussion above that these spectra (cone injected
slabs)
 have {\it strong} beaming along the {\bf B}-field. At lower fields the
beaming 
 is strongest.  On the other hand, note that fan injected slabs at lower
fields
 have the most isotropic emergent radiation field; this fact should make
their
 spectra easier to interpret.  The line wings are strongest in this case.
 An absorption-like line can be seen now for large viewing angles
 (i.e. the fundamental at $\mu < .25$).  These two
 observations also suggest redistribution out of the injected pattern and
into
 the solid angle space with less photon occupation relative to the 
 continuum.  The line wings for 
 $\mu \sim 1$ are exaggerated because the continuum is low there.
Alternatively,
 there is almost twice as much $\unvc_z$ flux bias in the isotropic
injection 
 case and the line wings at small angles are less prominent.  
 This lends further support to our proposal regarding the connection
 between angular redistribution and the formation of the line wings.

      Cone injected cylinders show strengthened signatures at the
 fundamental energy for $\mu>.25$, 
 compared to the isotropically injected case 
 (Fig.\ref{fig:Is.v.Cob.04}). 
 However, for small angles the line wings are depleted.
 Redistribution out of the cone accounts for the weakening
 of the line wings in cylinders as well.  
 Fewer photons are scattered back into the small angles than
 are scattered out.  The fan injected cylinders,
 Fig.\ref{fig:Is.v.Fab.04}, exhibit a reverse trend: 
 Line wings dominate except for $\mu \sim 0$ where
 there is virtually no feature at the fundamental for $B'$ = .04.
 Although this is noteworthy, 
 it is very unique to the parameter space of that particular model.

      Higher field plots show a general trend of broadening in all the
lines
 due to the higher temperatures required for detailed balance of cyclotron 
 cooling and heating rates (Lamb, Wang, Wasserman 1990).  For $B' = .24$, 
 the extremely broad line wings in the isotropic injection case overwhelm
the
 continuum at small angles in slabs.  Again these wings completely go away
when
 the photons are injected in a cone beam and an absorption-like line for 
 $\mu \sim 0$ appears as it did in lower fields.

 	Lastly, the higher field spectra exhibit a feature with strong 
 asymmetry at the fundamental energy.  This is a reflection of having 
 {\it step}-like profiles at large angles due to the 
 relativistic cut-off energy, in addition to
 lesser relative ratio between the profiles throughout the angle range. The
 profiles at lower fields show a `cusp' at this energy and thus, the
profile
 changes rapidly on either side of the cut-off energy.  
 The profile at higher fields shows no such cusp; 
 instead there is a sharp drop for energies beyond
 the relativistic cut off (Araya 1996, Araya \& Harding 1999).

\section{Conclusion}
 \label{sec:Con}

     This study has illustrated that the angular distribution of the
incident
 radiation has important effects on the emergent cyclotron line spectra.
 There is a profound interplay between the incident photon angular
distribution 
 and the geometry of the scattering region.  In general, the line features
are
 most pronounced along the directions of the anisotropic continuum
injection 
 and tend to be weakened in other directions, relative to the line features
 resulting
 from an isotropic continuum injection.  We find that the enhancements at
 the line wings of the 
 fundamental, which appear prominantly in the case of isotropic continuum 
 injection, are strongly suppressed along the direction of 
 anisotropy in the case
 of beamed continuum injection, regardless of geometry or beaming pattern. 

 The absence of line wings at the fundamental may
 thus be an indicator of non-isotropic continuum radiation. These trends
are 
 strong enough that 
 phase resolved spectral studies of observed cyclotron line features 
 of X-ray pulsars may be
 able to identify the continuum anisotropy orientation relative to the 
 direction of maximum optical depth of the scattering region.  It thus may
be  
 possible to obtain information about the continuum emission mechanism
 from the shapes of the observed cyclotron lines in these sources.

\acknowledgements
	We would like to thank Dr. R. Sina for supplying the code to 
 calculate the cross section using Sokolov-Ternov electron wavefunctions, 
 and Dr. A. Szalay for allowing us access to the supercomputer resources at
 the Johns Hopkins University.

%######################### Non-Isotropic Injection#########################

% ########################## Iso.v.Cone_b.04.ps ###########################
%--\figcaption{
\figureout{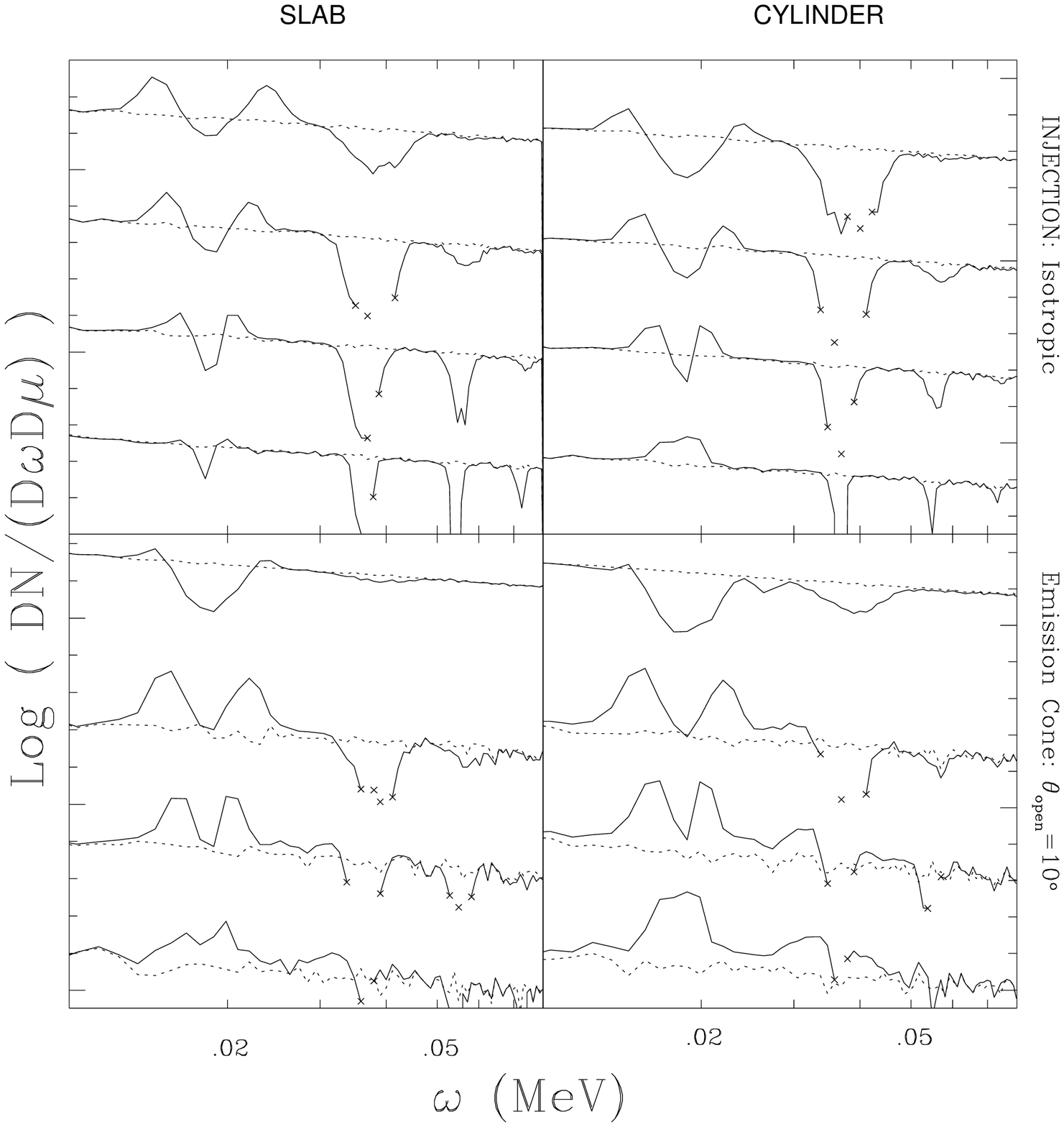}{6.0in}{0}
{{\bf Comparison between cone and isotropic injection for slab and 
   cylindrical geometries.}  The magnetic field strength is $B'$ = .04
(1.7TG).
   Each plot shows angle dependent model photon spectra. On each quadrant, 
   the bottom plot is the spectrum emerging from $\mu < .25$, followed by
   spectra from $.25 < \mu < .50$, from $.5 < \mu < .75$ and from $\mu
>.75$
   for the top plot. Each run results from 50 thousand photons injected
   isotropically, for the top plots, or within a lorentzian cone with
   opening angle $\theta_{op} = 10^o$ for the bottom plots.
   {\bf Dotted line}: injected continuum: power law with $\alpha = -1$.
   {\bf Solid line}: output scattered spectrum.
   {\bf Crosses}: areas where photon depletion has occurred, the model
points 
                  are indicated as crosses for ease of reading.
   $\tau_c = 1 \times 10^{-3}, ~~T_e = \fourth \omega_{cyc}$
   The flux normalization is arbitrary. 
   These are local spectra emitted near the neutron star surface and do not
   include General relativistic effects such as red shift or light bending.
   \label{fig:Is.v.Cob.04} 
   }

% ########################## AngDistb.04C.ps  ###########################

\figureout{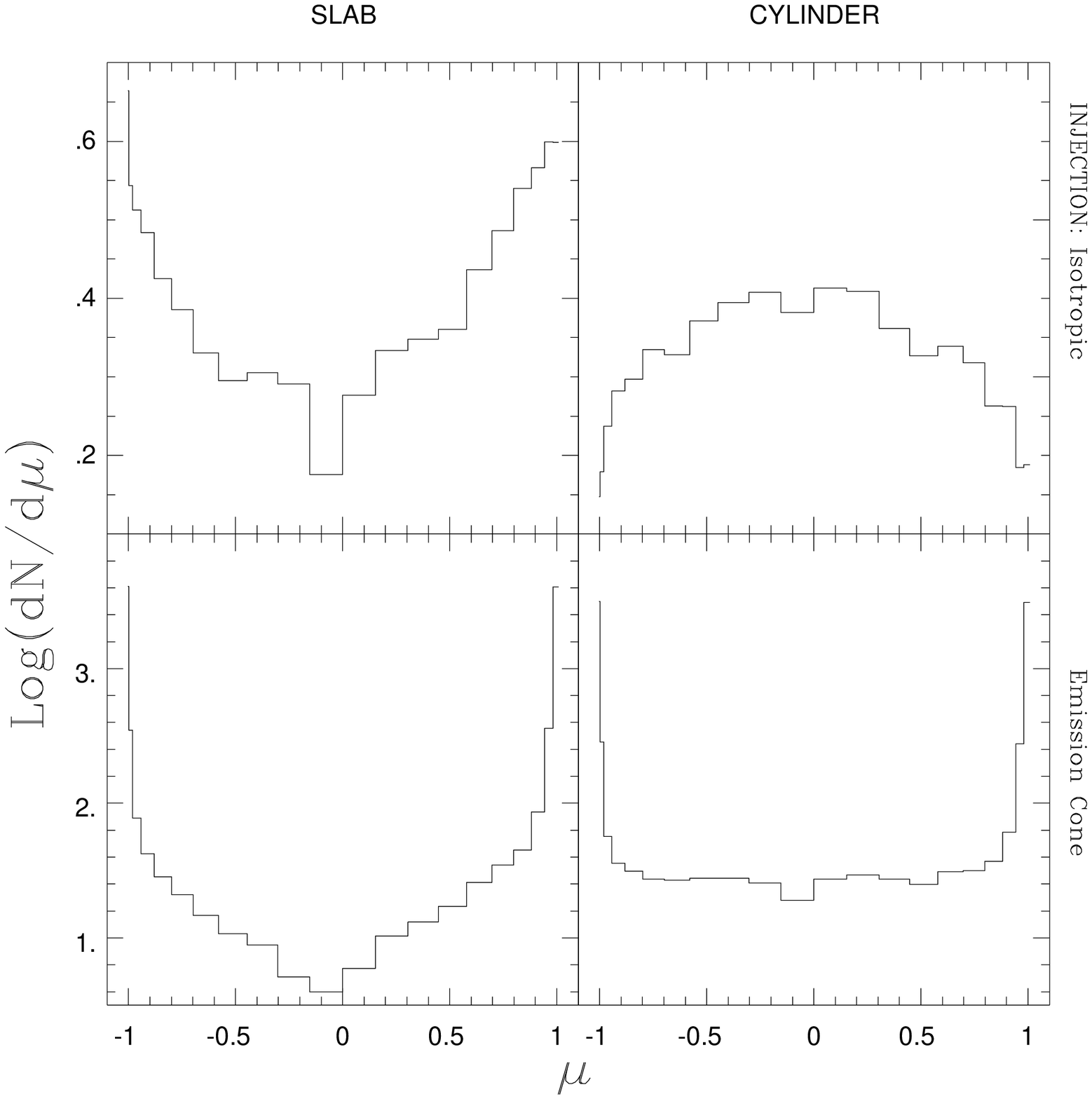}{6.0in}{0}
   {{\bf Comparison of emergent angular distributions for cone and isotropic
   injections in slab and cylindrical geometries.}
   The parameters are identical with those of the preceding figure.
   \label{fig:AngDistb.04C} 
   }

% ########################### Iso.v.Fan_b.04.ps #########################
\figureout{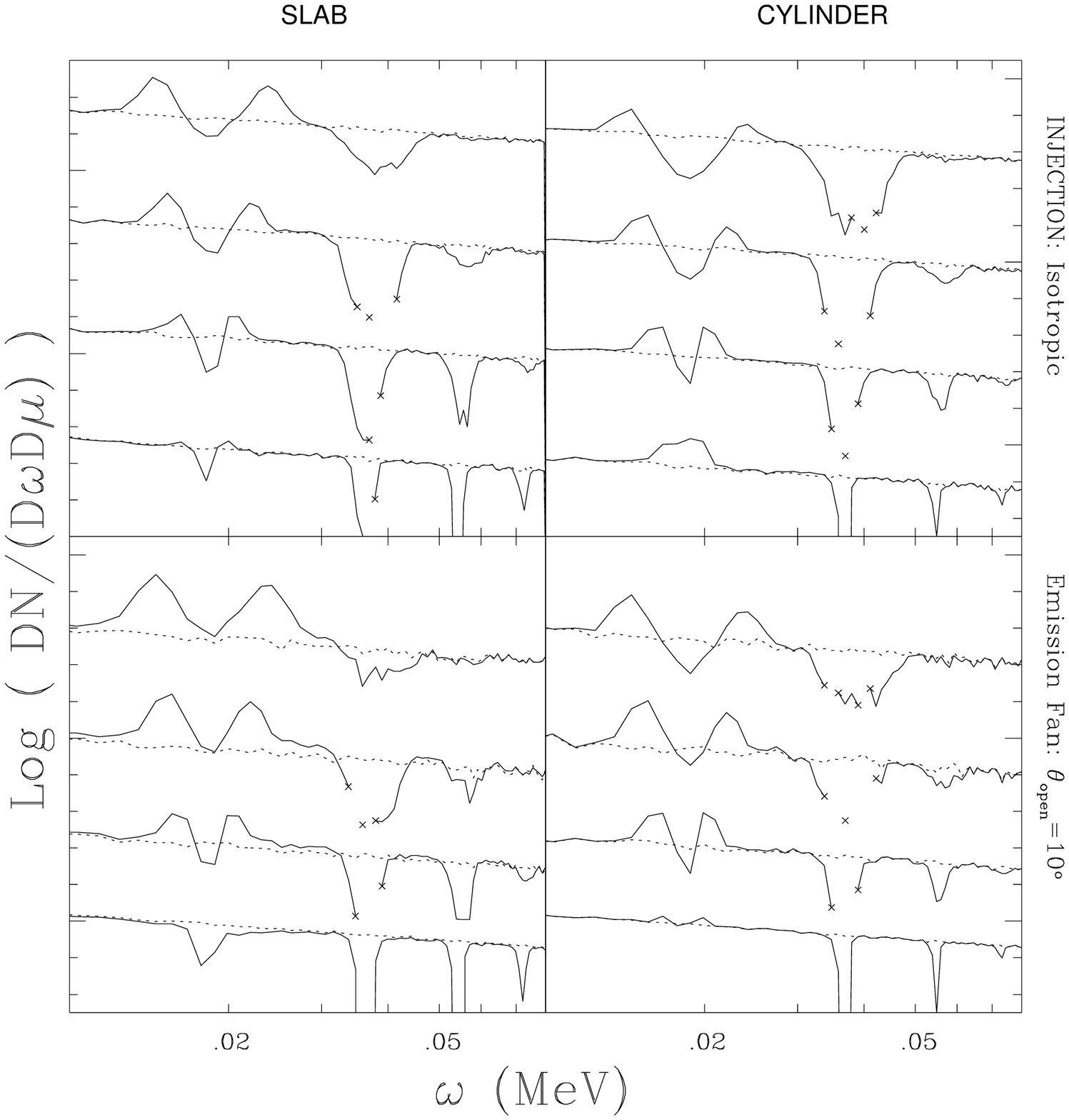}{6.0in}{0}
{{\bf Comparison between fan and isotropic injection for slab and 
   cylindrical geometries.}  The magnetic field strength is $B'$ = .04 (1.7
TG).
   Each plot shows angle dependent model photon spectra. On each quadrant, 
   the bottom plot is the spectrum emerging from $\mu < .25$, and the top
plot
   shows the spectrum from $\mu > .75$, with similar binning in between.
   Each run results from 50 thousand photons injected isotropically, for
the
   top plots, or within a lorentzian fan (injection perpendicular to {\bf
B})
   with opening angle $\theta_{op} = 10^o$, for the bottom plots.
   {\bf Dotted line}: injected continuum: power law with $\alpha = -1$.
   {\bf Solid line}: output scattered spectrum.
   {\bf Crosses}: areas where photon depletion has occurred, the model
points 
                  are indicated as crosses for ease of reading.
   $\tau_c = 1 \times 10^{-3}, ~~T_e = \fourth \omega_{cyc}$
   The flux normalization is arbitrary. 
   \label{fig:Is.v.Fab.04}
   }

% ########################## AngDistb.04F.ps  ###########################
\figureout{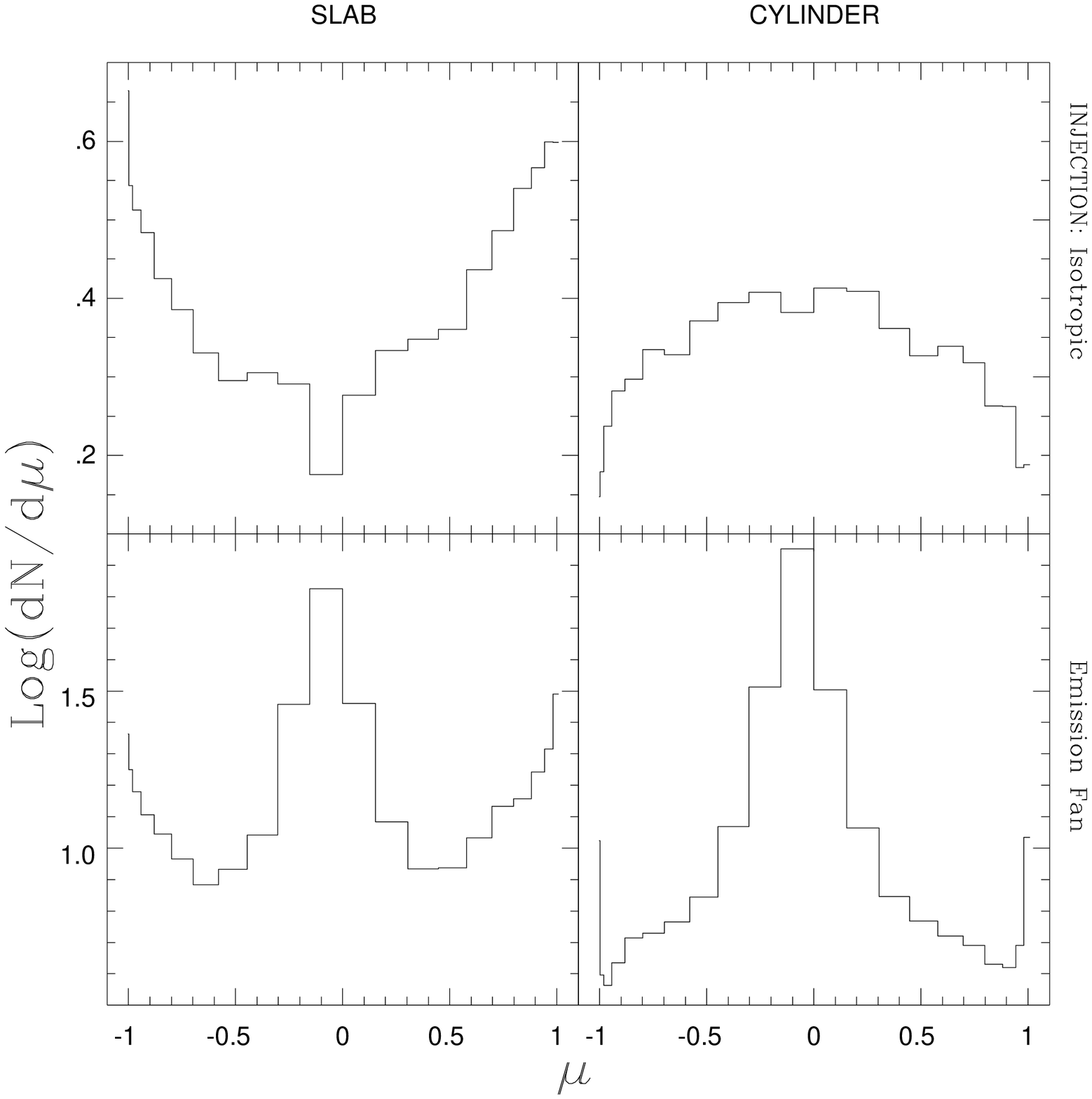}{6.0in}{0}
   {{\bf Comparison of emergent angular distributions for fan and isotropic
   injections in slab and cylindrical geometries.}
   The parameters are identical with those of the preceding figure.
   \label{fig:AngDistb.04F} 
   }

% ########################## Iso.v.Cone_b.1.ps  ###########################
\figureout{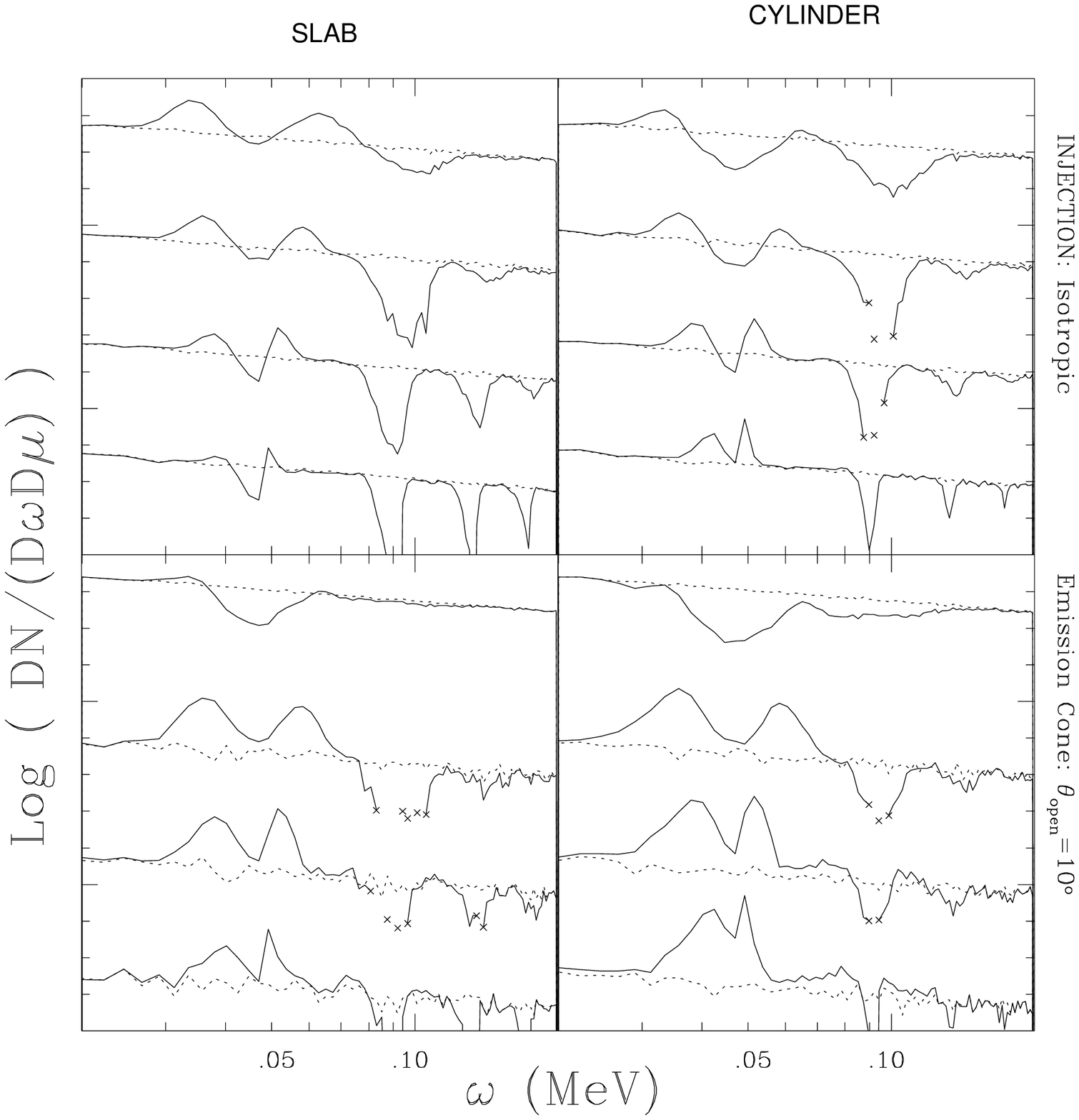}{6.0in}{0}
   {{\bf Comparison between cone and isotropic injection for slab and
   cylindrical geometries.} The magnetic field strength is $B'$ = .1 (4.414
TG).
   Each plot shows angle dependent model photon spectra. On each quadrant, 
   the bottom plot is the spectrum emerging from $\mu < .25$, and the top
plot
   shows the spectrum from $\mu > .75$, with similar binning in between.
   Each run has 50 thousand photons. Top plots correspond to isotropic
   injection, bottom plots to beamed injection, as previously described.
   {\bf Dotted line}: injected continuum: power law with $\alpha = -1$.
   {\bf Solid line}: output scattered spectrum.
   {\bf Crosses}: areas where photon depletion has occurred.
   $\tau_c = 1 \times 10^{-3}, ~~T_e = \fourth \omega_{cyc}$
   The flux normalization is arbitrary. 
   \label{fig:Is.v.Cob.1}
   }

% ########################## AngDistb.10C.ps  ##########################
\figureout{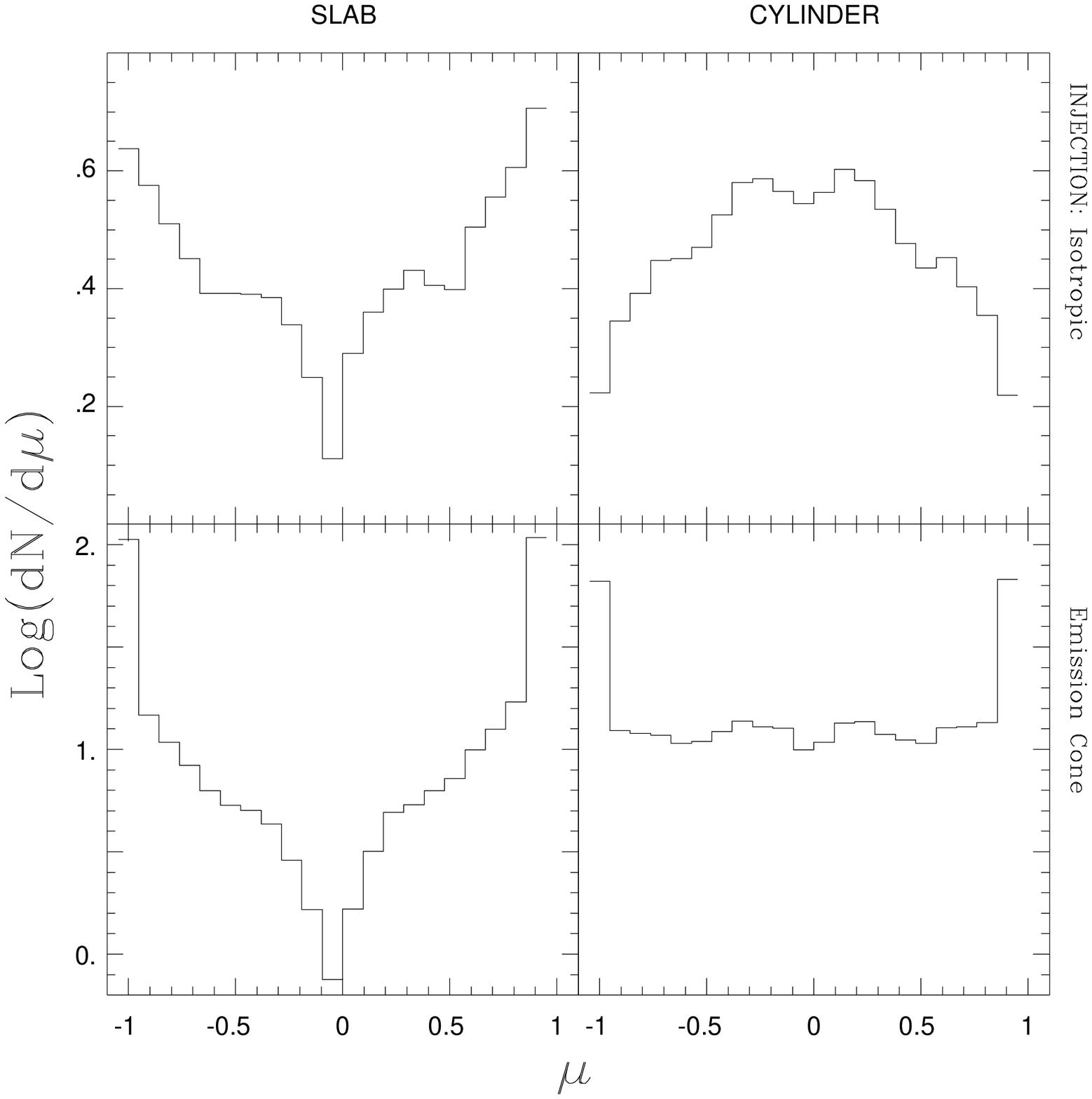}{6.0in}{0}
   {{\bf Angular distribution comparison between cone and isotropic
injection
   for slab and cylindrical geometries.}
   The parameters are identical with those of the preceding figure.
   \label{fig:AngDistb.10C} 
   }

% ########################## Iso.v.Fan.1.ps  ###########################
\figureout{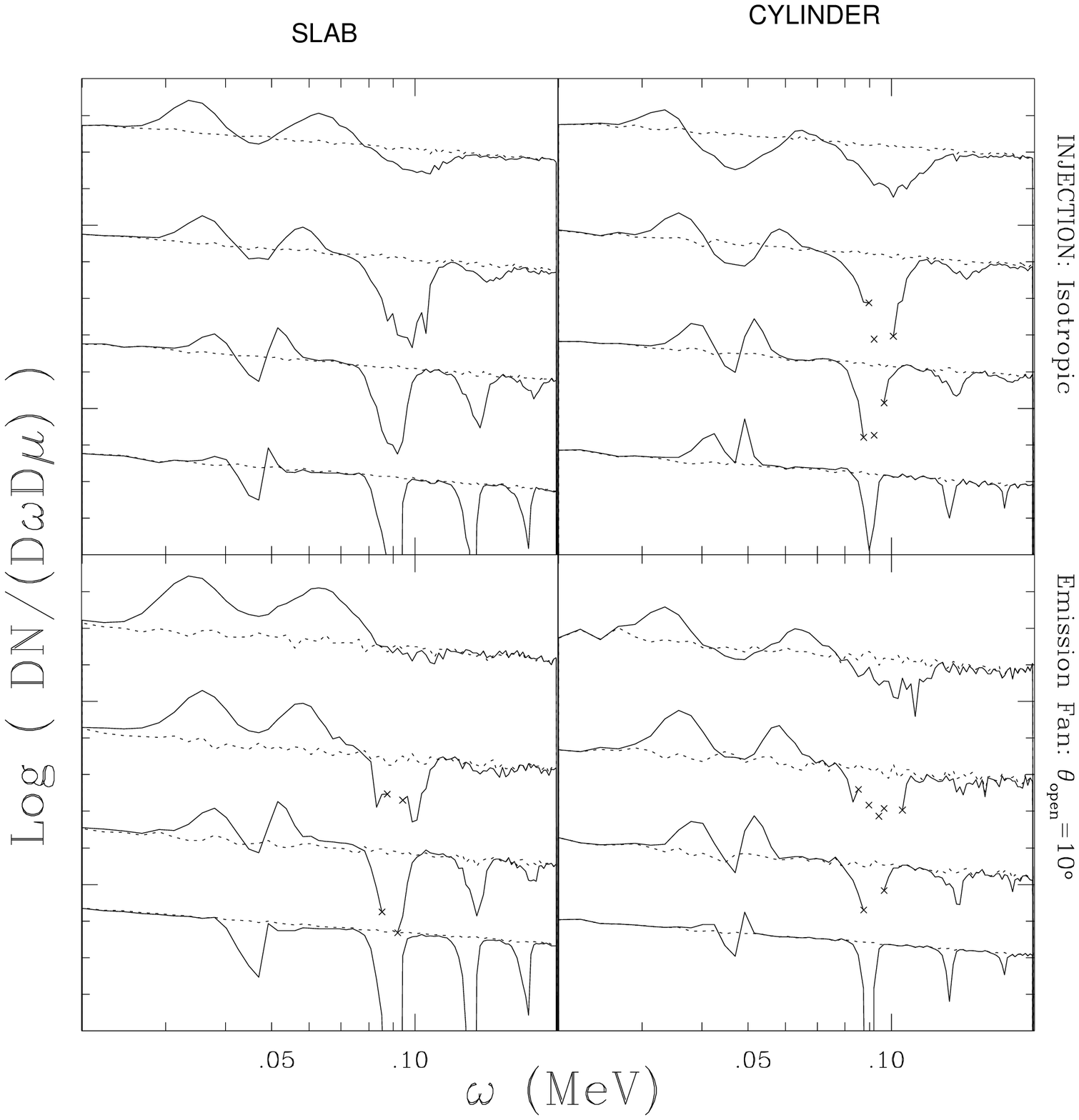}{6.0in}{0}
   {{\bf	 Comparison between fan and isotropic injection for slab and 
   cylindrical geometries.} The magnetic field strength is $B'$ = .1 (4.414
TG).
   Each plot shows angle dependent model photon spectra. On each quadrant, 
   the bottom plot is the spectrum emerging from $\mu < .25$, and the top
plot
   shows the spectrum from $\mu > .75$, with similar binning in between.
   Each run has 50 thousand photons. Top plots correspond to isotropic
   injection, bottom plots to beamed injection, as previously described.
   {\bf Dotted line}: injected continuum: power law with $\alpha = -1$.
   {\bf Solid line}: output scattered spectrum.
   {\bf Crosses}: areas where photon depletion has occurred.
   $\tau_c = 1 \times 10^{-3}, ~~T_e = \fourth \omega_{cyc}$
   The flux normalization is arbitrary. 
   \label{fig:Is.v.Fab.1}
  }

% ########################## AngDistb.10F.ps  ###########################
\figureout{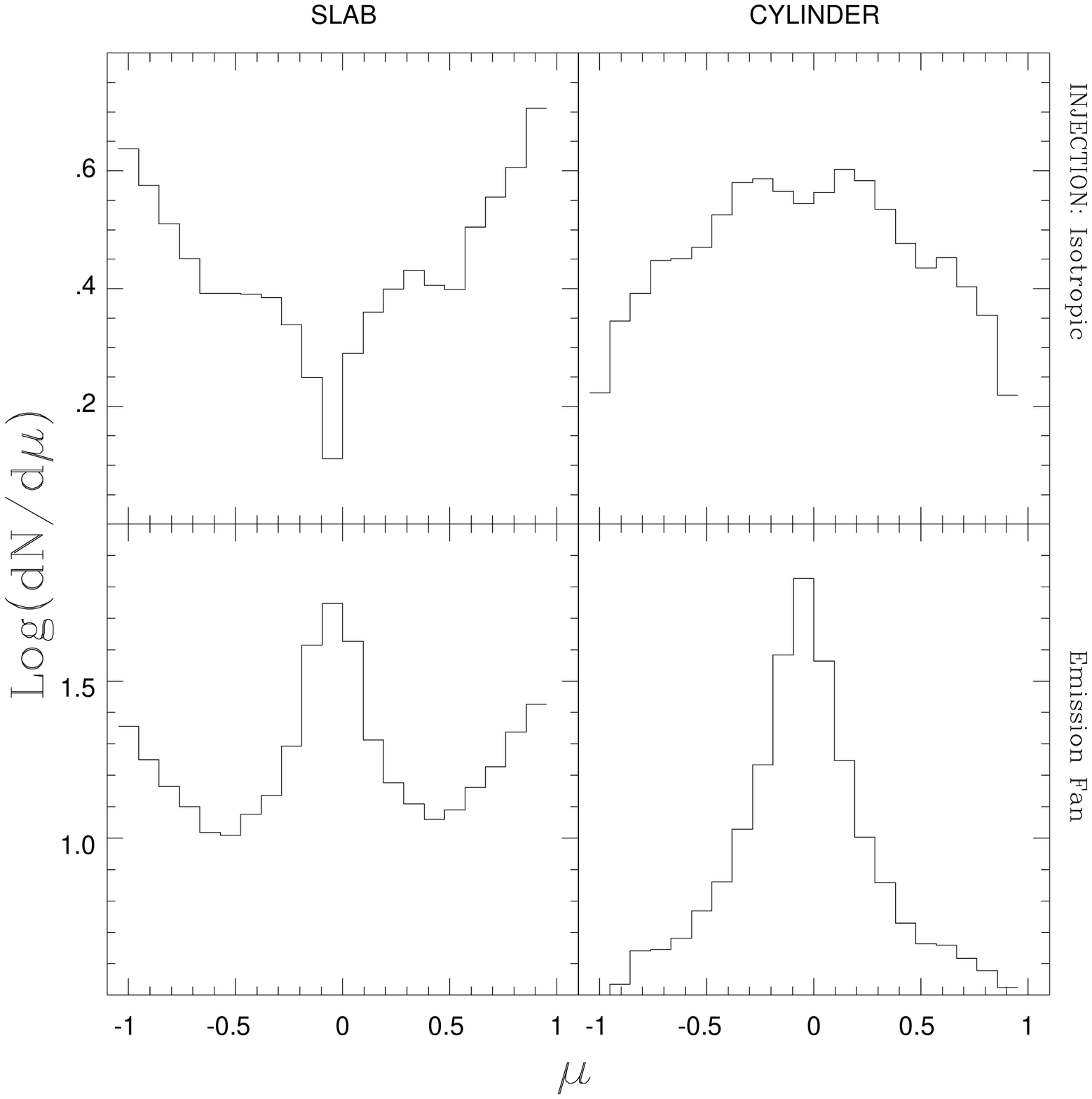}{6.0in}{0}
   {{\bf Comparison of emergent angular distributions for fan and isotropic
   injections in slab and cylindrical geometries.}
   The parameters are identical with those of the preceding figure.
   \label{fig:AngDistb.10F} 
   }

% ########################## Iso.v.Cone_b.24.ps ###########################
\figureout{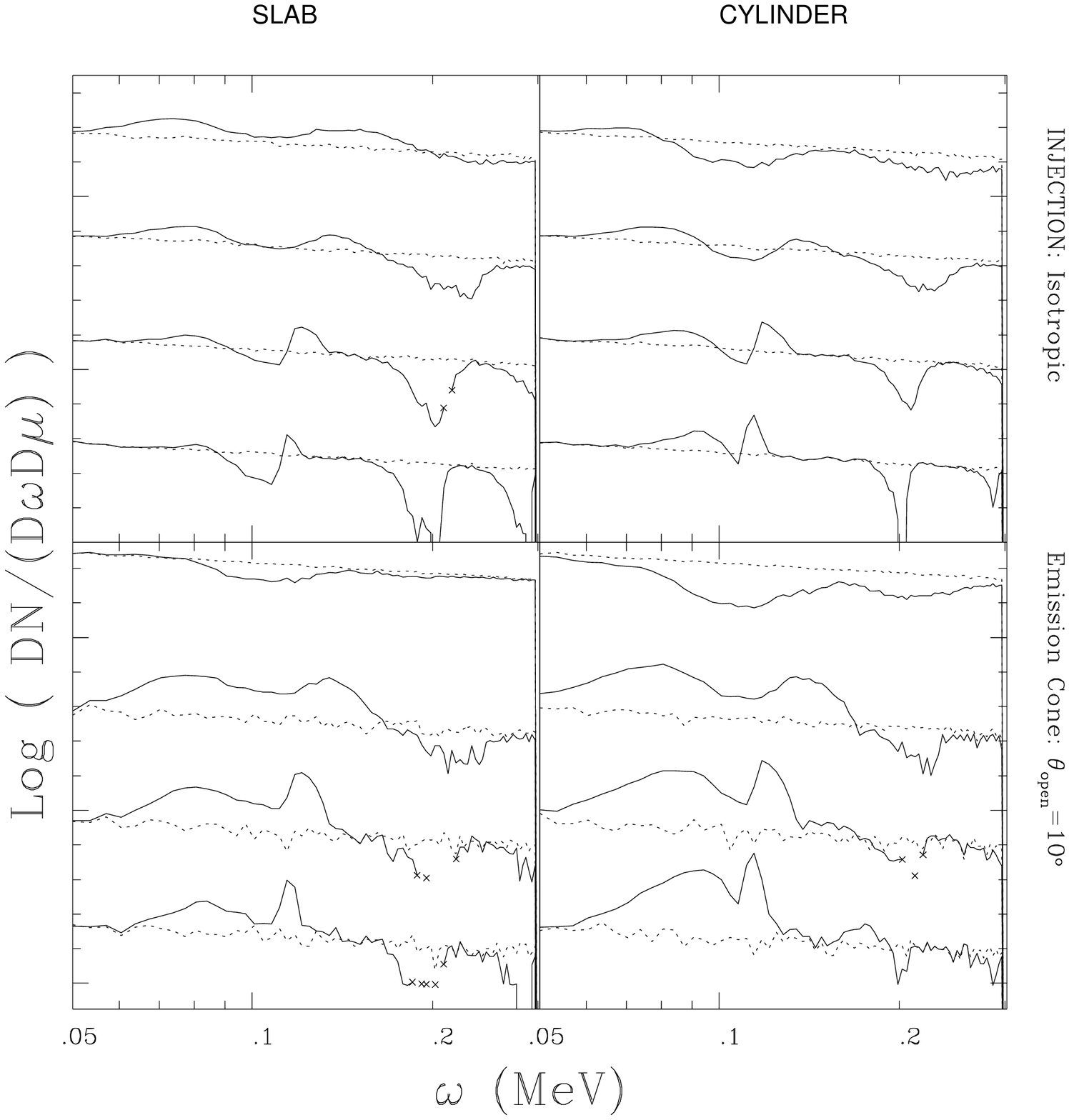}{6.0in}{0}
   {{\bf  Comparison between cone and isotropic injection for slab and 
   cylindrical geometries.}  The magnetic field strength is $B'$ = .2436 
   (10.75 TG).  Each plot shows angle dependent model photon spectra. 
   On each quadrant, the bottom plot is the spectrum emerging from
   $\mu < .25$, and the top plot shows the spectrum from $\mu > .75$,
   with similar binning in between.  Each run has 50 thousand photons. 
   Top plots correspond to isotropic
   injection, bottom plots to beamed injection, as previously described.
   {\bf Dotted line}: injected continuum: power law with $\alpha = -1$.
   {\bf Solid line}: output scattered spectrum.
   {\bf Crosses}: areas where photon depletion has occurred.
   $\tau_c = 1 \times 10^{-3}, ~~T_e = \fourth \omega_{cyc}$
   The flux normalization is arbitrary. 
   \label{fig:Is.v.Cob.24} 
   }

% ########################## AngDistb.24C.ps  ###########################
\figureout{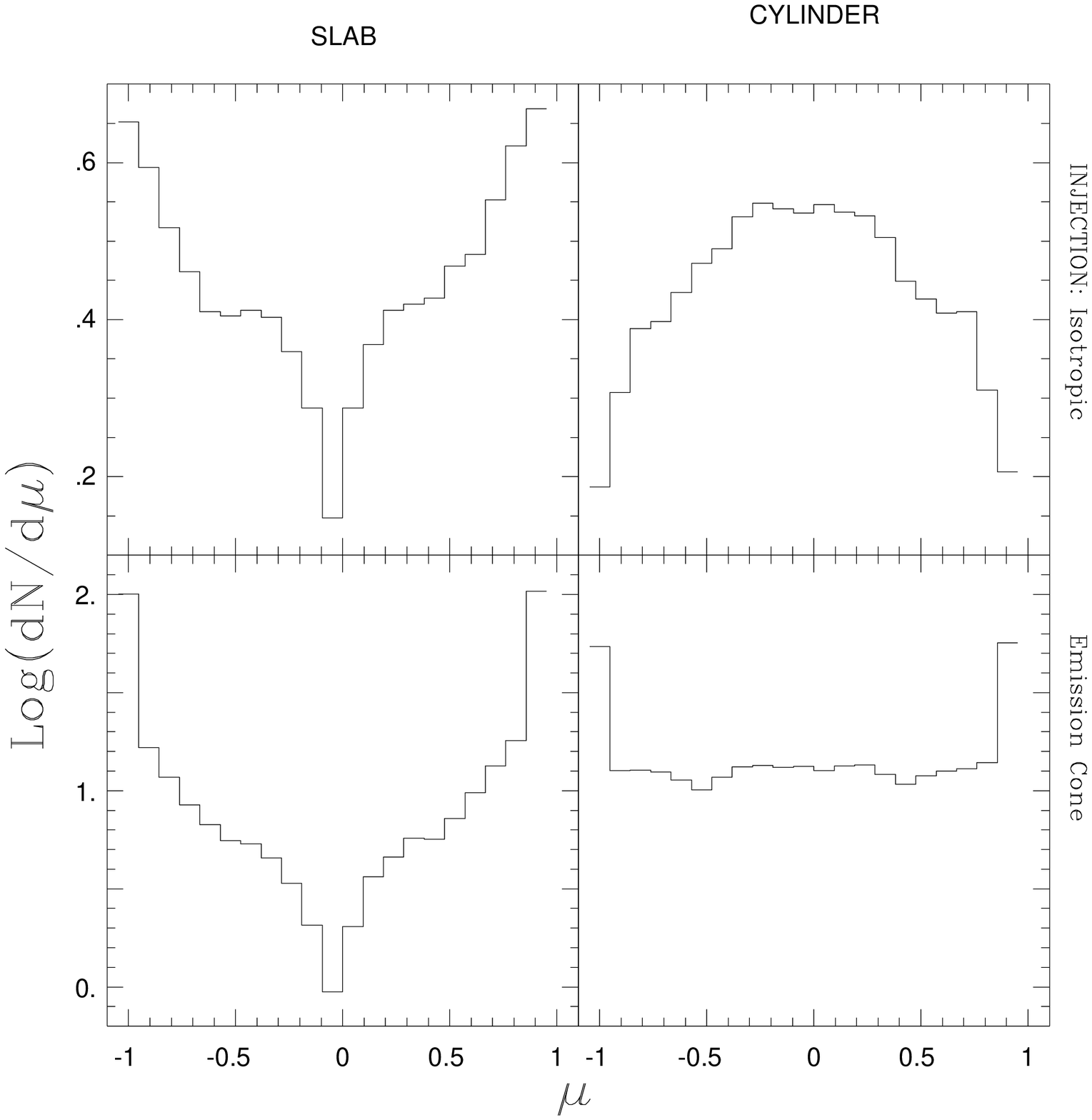}{6.0in}{0}
   {{\bf Comparison of emergent angular distributions for cone and isotropic
   injections in slab and cylindrical geometries.}
   The parameters are identical with those of the preceding figure.
   \label{fig:AngDistb.24C} 
   }

% ########################## Iso.v.Fan_b.24.ps  ###########################
\figureout{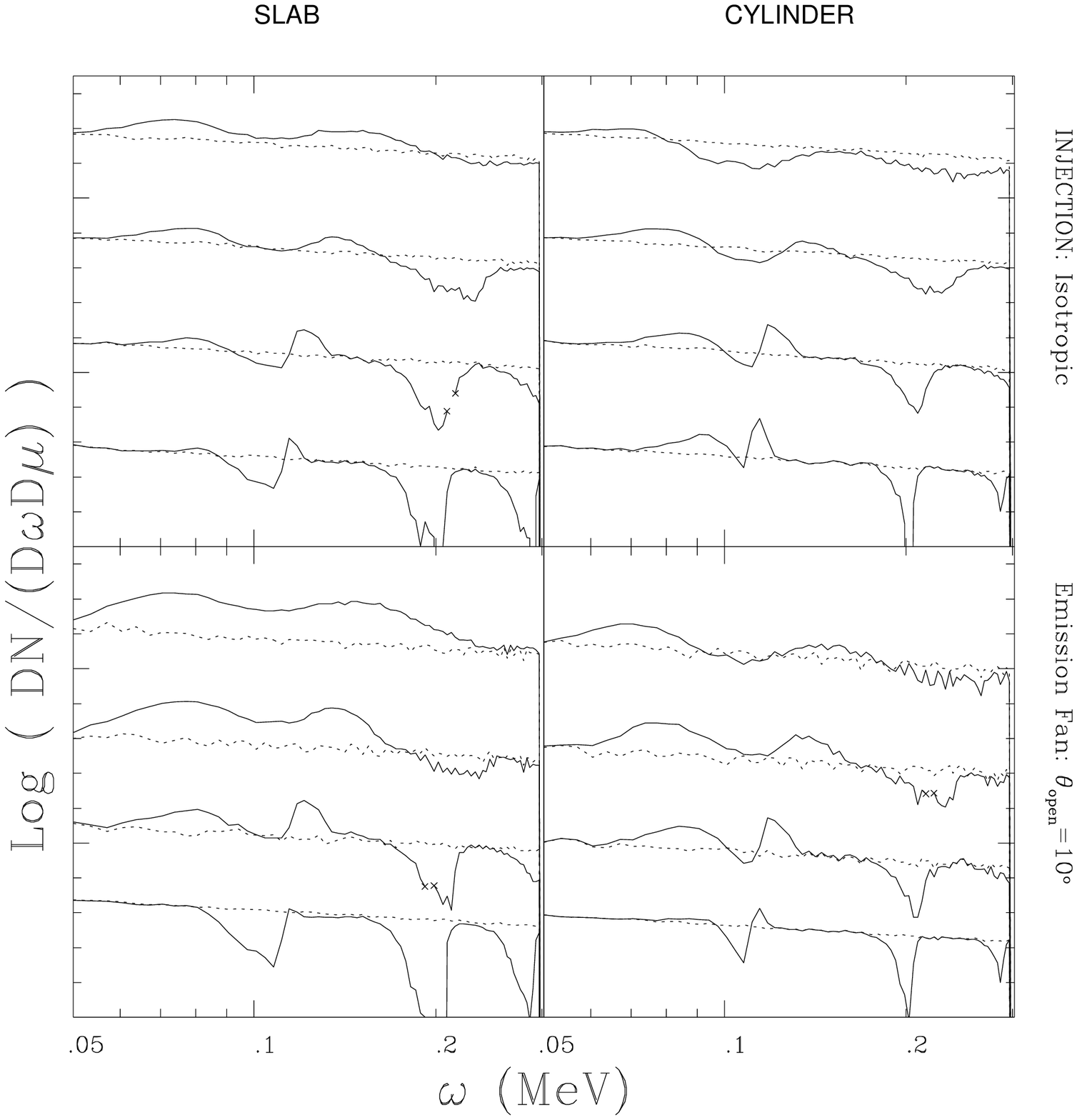}{6.0in}{0}
   {{\bf	 Comparison between fan and isotropic injection for slab and 
   cylindrical geometries.}  The magnetic field strength is $B'$ = .2436 
   (10.75 TG).  Each plot shows angle dependent model photon spectra. 
   On each quadrant, the bottom plot is the spectrum emerging from
   $\mu < .25$, and the top plot shows the spectrum from $\mu > .75$,
   with similar binning in between.  Each run has 50 thousand photons. 
   Top plots correspond to isotropic
   injection, bottom plots to beamed injection, as previously described.
   {\bf Dotted line}: injected continuum: power law with $\alpha = -1$.
   {\bf Solid line}: output scattered spectrum.
   {\bf Crosses}: areas where photon depletion has occurred.
   $\tau_c = 1 \times 10^{-3}, ~~T_e = \fourth \omega_{cyc}$
   The flux normalization is arbitrary. 
   \label{fig:Is.v.Fab.24}
   }

% ########################## AngDistb.24F.ps  ###########################
\figureout{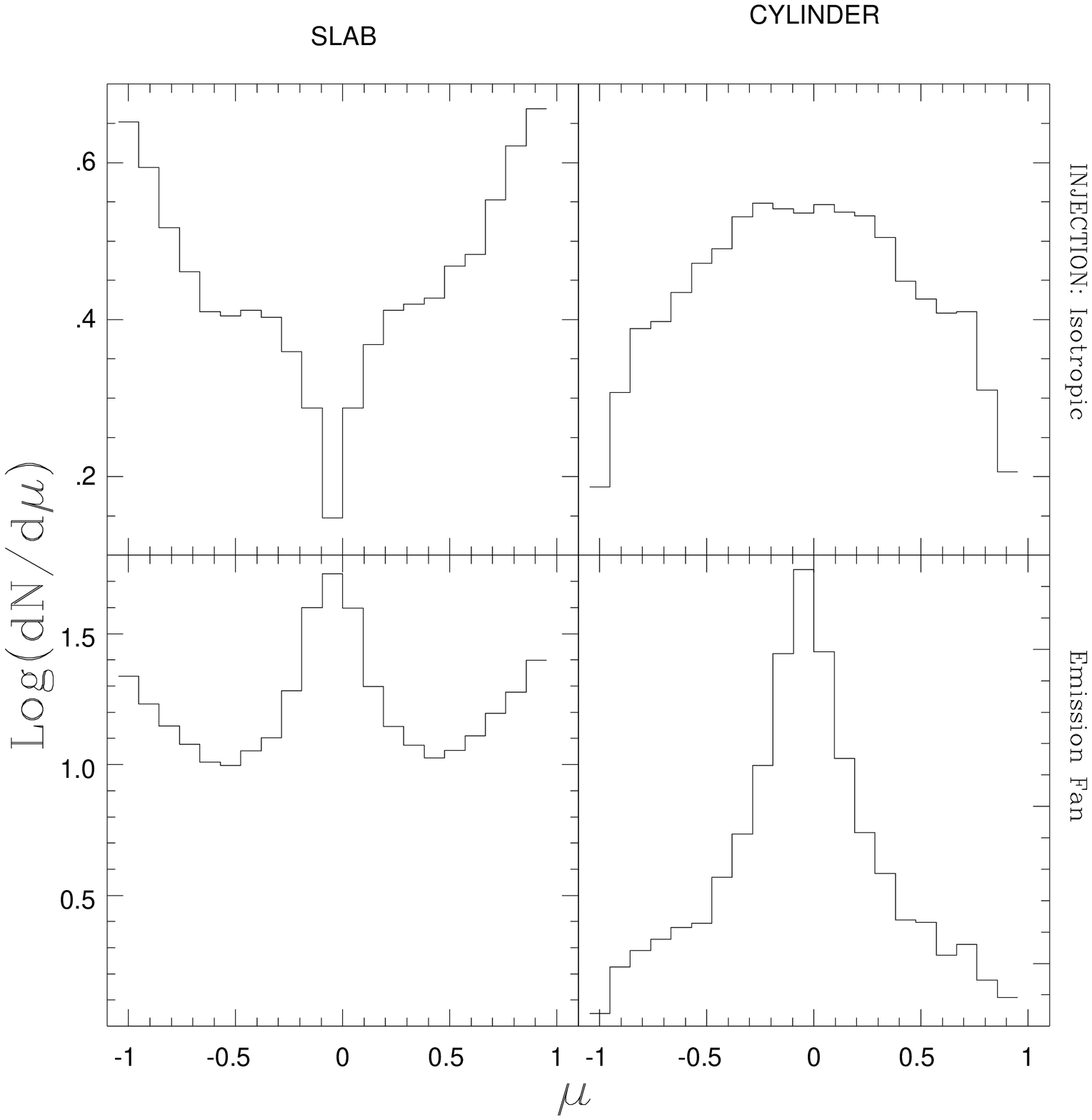}{6.0in}{0}
   {{\bf Comparison of emergent angular distributions for fan and isotropic
   injections in slab and cylindrical geometries.}
   The parameters are identical with those of the preceding figure.
   \label{fig:AngDistb.24F} 
   }
\end{document}